\documentclass[letterpaper,10pt]{article}

\usepackage{amssymb}
\usepackage{amsmath}

\newcommand{\dd}{\textrm{d}}

\author{A. L\'opez-Ortega\thanks{alopezo@ipn.mx} \\
Centro de Investigaci\'on en Ciencia Aplicada y Tecnolog\'{\i}a Avanzada. \\ 
Unidad Legaria. Instituto Polit\'ecnico Nacional. \\
Calzada Legaria \# 694. Colonia Irrigaci\'on. Delegaci\'on Miguel Hidalgo. \\
M\'exico, D.\ F., M\'exico. \\
C.\ P.\  11500  
}

\title{Entropy spectrum of the $D$-dimensional massless topological black hole}

\begin{document}

\maketitle

\begin{abstract}

There are exact solutions to Einstein's equations with negative cosmological constant that represent black holes whose event horizons are manifolds of negative curvature, the so-called topological black holes. Among these solutions there is one, the  massless topological black hole, whose mass is equal to zero. Hod proposes that in the semiclassical limit the asymptotic quasinormal frequencies determine the entropy spectrum of the black holes. Taking into account this proposal, we calculate the entropy spectrum of the massless topological black hole and we compare with the results on the entropy spectra of other topological black holes.



\end{abstract}

Bekenstein proposes that in a quantum theory of gravity the horizon area of a black hole is quantized \cite{Bekenstein:1974jk}, \cite{Bekenstein:1973ur}. Also in the semiclassical limit he expects that the area spectrum takes the form
\begin{equation}
 A_n \approx \epsilon \hbar n ,
\end{equation} 
where $n=0,1,2,\dots$, $\hbar$ denotes the reduced Planck constant, and $\epsilon$ stands for an unknown parameter of order 1.

In an interesting conjecture \cite{Hod:1998vk}, Hod states that the spacing of the area spectrum ($\epsilon \hbar$) is determined by the real part of the asymptotic quasinormal frequencies (QNF) of the black hole under study. Based on Hod's ideas Kunstatter proposes a different method to obtain equivalent results for the area spectrum \cite{Kunstatter:2002pj}. Hod's conjecture gives reasonable results for several black holes \cite{Hod:1998vk}, \cite{Kunstatter:2002pj}, but for other black holes it leads to confusing results, as for example \cite{Maggiore:2007nq}, \cite{Setare:2004uu}:
\begin{enumerate}

\item Based on the results of other methods, there are black holes for which in the semiclassical limit we expect that their area quanta be equal, but the real parts of their asymptotic QNF are not equal, hence Hod's conjecture predicts different values for the area quanta.

\item For some black holes we expect a discrete and evenly spaced area spectrum, nevertheless Hod's conjecture predicts a discrete  but not evenly spaced area spectrum  \cite{Setare:2004uu}.

\end{enumerate}

To solve some of these problems, Maggiore \cite{Maggiore:2007nq} suggests that the quasinormal modes of a black hole behave as the oscillations of a damped oscillator whose frequencies of oscillation are $\omega_{p,k}=\sqrt{\omega_R^2 + \omega_I^2}$ (the so-called physical frequencies), where $\omega_R$ and $\omega_I$ stand for the real and imaginary parts of the QNF. Moreover, based on Hod's conjecture, Maggiore proposes that the area spectrum of the black hole is determined by the asymptotic value of the physical frequencies \cite{Maggiore:2007nq}. Recently Maggiore's proposal has been used to calculate the area and entropy spectra of several black holes \cite{Maggiore:2007nq}, \cite{Wei:2009yj}--\cite{LopezOrtega:2010tg}.

According to Birmingham and Mokhtari \cite{Birmingham:2006zx} in several lines of research the so-called $D$-dimensional ($D \geq 4$) massless topological black hole (MTBH) may play a similar role to the three-dimensional BTZ black hole \cite{Banados:1992wn}. The MTBH is a static solution to Einstein's equations with negative cosmological constant, whose mass is equal to zero and its event horizon is a compact and orientable manifold of negative curvature  
\cite{Mann:1997iz}--\cite{Lemos:1994xp}. The MTBH has been studied thoroughly and we note that this black hole is member of the exact solutions usually known as topological black holes whose event horizons are  manifolds of negative curvature \cite{Mann:1997iz}--\cite{Lemos:1994xp}. It is convenient to notice that in gravity theories different from general relativity there are exact solutions that represent topological black holes (for example see \cite{Gonzalez:2010vv}, \cite{Klemm:1998kf}--\cite{Aros:2000ij}, \cite{Oliva:2010xn} for more details).

Taking into account the ideas by Hod, Kunstatter, and Maggiore \cite{Hod:1998vk}--\cite{Maggiore:2007nq} in what follows we calculate the area and entropy spectra of the MTBH. Recently Gonzalez et al.\ \cite{Gonzalez:2010vv} compute the entropy and mass spectra of more general topological black holes (Chern-Simons black holes). Since it is difficult to identify the entropy spectrum of the MTBH in the results of \cite{Gonzalez:2010vv}, and also to know its entropy spectrum is interesting in itself, it is appropriate to calculate the entropy spectrum of the MTBH and compare with the entropy spectra of other topological black holes.  Furthermore we make some comments on the results of \cite{Gonzalez:2010vv}.

The metric of the $D$-dimensional MTBH reads
\begin{equation} \label{eq: MTBH metric}
\dd s^2 = -\left(-1 + \frac{r^2}{L^2} \right) \dd t^2 + \left(-1 + \frac{r^2}{L^2} \right)^{-1} \dd r^2 + r^2 \dd \Sigma^2_{D-2},
\end{equation} 
where $r\in (L, +\infty)$, $L$ is related to negative cosmological constant of the spacetime, and $\dd \Sigma^2_{D-2}$ stands for the line element of a $(D-2)$-dimensional compact and orientable Einstein manifold of negative curvature $\Sigma_{D-2}$ \cite{Mann:1997iz}--\cite{Lemos:1994xp}.

The QNF of the $D$-dimensional MTBH (\ref{eq: MTBH metric}) are calculated exactly for the Klein-Gordon \cite{Aros:2002te}, \cite{Oliva:2010xn}, electromagnetic \cite{Oliva:2010xn}, \cite{LopezOrtega:2007vu},  Dirac \cite{LopezOrtega:2010}, and gravitational \cite{Birmingham:2006zx} perturbations. In particular the QNF of the gravitational perturbations are (see formulas (46) of \cite{Birmingham:2006zx})
\begin{equation} \label{eq: GP MTBH}
 \omega =\pm \frac{\xi}{L} - \frac{2i}{L}\left( k + \frac{\mathbb{A}}{4} \right) ,
\end{equation} 
where $k=0,1,2,\dots$, the parameter $\mathbb{A}$ is equal to
 \begin{align} \label{eq: A values MTBH}
 \mathbb{A}=  & \left\{ \begin{array}{l}  D-1 \qquad \, \, \, \, \, \textrm{for the vector type} \\ 
\quad \qquad \,\, \, \, \, \, \,\, \quad \textrm{gravitational perturbations}, \\
 |D-5|+2  \, \, \, \,\textrm{for the scalar type} \\
\quad \qquad \,\, \, \, \,\, \quad \,\,\,\textrm{gravitational perturbations}, \\
 D+1  \, \,\, \,\, \, \qquad \textrm{for the tensor type} \\
\quad \quad \,\, \, \, \,\, \qquad \,\, \, \textrm{gravitational perturbations ($D\geq5$),}
 \end{array} \right.
\end{align}   
 and the quantities $\xi$ are the eigenvalues of the scalar, vector, and tensor harmonics on the base manifold $\Sigma_{D-2}$. Thus the quantities  $\xi$ depend on the type and the mode of the gravitational perturbations.

Since the real part of the QNF ($\xi/L$) depends on the eigenvalues of the harmonics on the base manifold $\Sigma_{D-2}$, we believe that the original Hod's conjecture does not hold for the MTBH. To see this fact we use the original Hod's conjecture with the real part of the QNF (\ref{eq: GP MTBH}) to get the area spectrum $A_n=8\pi \hbar \xi n$ for the horizon of the $D$-dimensional MTBH and we note that it depends on the type and the mode of the gravitational perturbations that we use in the calculation.

In the asymptotic limit $k \to \infty$ the QNF of the gravitational perturbations (\ref{eq: GP MTBH}) reduce to
\begin{equation}
 \omega \approx - \frac{2 i}{L} k.
\end{equation} 
(A similar expression is valid for the QNF of other fields.) Therefore we find that the asymptotic physical frequencies of the MTBH are
\begin{equation}
 \omega_{p,k}=\frac{2k}{L}.
\end{equation}  

Kunstatter \cite{Kunstatter:2002pj} points out that for a system of energy $E$ and oscillation frequency $\omega$, the integral
\begin{equation}
 I = \int \frac{\dd E}{\omega}
\end{equation} 
is an adiabatic invariant. In the framework of Maggiore's proposal \cite{Maggiore:2007nq} the change in the physical parameters of the black hole stems from a transition in the physical frequencies of type $\omega_{p,k+1} \to \omega_{p,k}$; hence for the black hole the appropriate frequency of oscillation is $\Delta \omega = \omega_{p,k+1} - \omega_{p,k}$ \cite{Vagenas:2008yi}, and the adiabatic invariant corresponding to the black hole is \cite{Kunstatter:2002pj}, \cite{Vagenas:2008yi}, 
\begin{equation}
 I = \int \frac{\dd M}{\Delta \omega}.
\end{equation} 

For the uncharged topological black holes of Einstein's theory the first law of the thermodynamics reads \cite{Vanzo:1997gw}--\cite{Wald:1999vt}
\begin{equation} \label{eq: first law}
 \dd M = \frac{\kappa }{8 \pi} \dd A ,
\end{equation} 
where $M$ stands for the mass of the black hole, $\kappa$ denotes the surface gravity of the event horizon, and $A$ stands for the horizon area. Taking into account the first law (\ref{eq: first law}) we find that for the MTBH the adiabatic invariant $I$ is
\begin{equation} \label{eq: MTBH adiabatic invariant}
 I = \frac{L}{2} \int \dd M = \frac{L}{2} \int \frac{\kappa}{8 \pi} \dd A = \frac{A}{16 \pi}.
\end{equation} 

In the semiclassical limit, by Bohr-Sommerfeld quantization rule, an adiabatic invariant is quantized in equal steps, thus $I \approx n \hbar$, and therefore from formula (\ref{eq: MTBH adiabatic invariant}) we get that in the semiclassical limit the area spectrum of the $D$-dimensional MTBH is
\begin{equation} \label{eq: area spectrum}
 A_n = 16 \pi \hbar n.
\end{equation} 
For the topological black holes of Einstein's gravity the Bekenstein-Hawking area-entropy law $S=A / 4 \hbar$ is valid \cite{Wald:1999vt}, \cite{Dias:2006vy}, therefore from the area spectrum (\ref{eq: area spectrum}) we obtain that the entropy spectrum of the MTBH is
\begin{equation} \label{eq: entropy spectrum}
 S_n = 4 \pi n.
\end{equation} 
Thus from the area spectrum (\ref{eq: area spectrum}) and entropy spectrum (\ref{eq: entropy spectrum}) we find that the area and entropy quanta are
\begin{equation} \label{eq: area entropy quanta}
 \Delta A = 16 \pi \hbar, \qquad \qquad \Delta S = 4 \pi .
\end{equation} 

For the $D$-dimensional MTBH their area and entropy quanta (\ref{eq: area entropy quanta}) are independent of the cosmological constant and the spacetime dimension. Furthermore the entropy quantum (\ref{eq: area entropy quanta}) does not depend on the properties of the base manifold $\Sigma_{D-2}$ (as its genus number or its volume). We must note this property of the entropy quantum since the MTBH is among the simplest topological black holes and this property is not valid for the topological black holes studied in \cite{Gonzalez:2010vv} (see below). 

The values (\ref{eq: area entropy quanta}) for the area and entropy quanta of the MTBH are equal to those obtained for the $D$-dimensional de Sitter spacetime \cite{LopezOrtega:2009ww}. Also the area and entropy quanta (\ref{eq: area entropy quanta}) are different from the quanta $ \Delta A = 8 \pi \hbar$ and $\Delta S = 2 \pi$ corresponding to the $D$-dimensional Schwarzschild black hole \cite{Maggiore:2007nq}, \cite{Wei:2009yj}, \cite{LopezOrtega:2010tg}, the $D$-dimensional Reissner-Nordstr\"om black hole in the small charge limit \cite{LopezOrtega:2010tg}, and the four-dimensional slowly rotating Kerr black hole \cite{Vagenas:2008yi}, \cite{Medved:2008iq}.

Daghigh and Green \cite{Daghigh:2008jz} calculate the entropy quanta for the $D$-dimensional Schwarzschild anti-de Sitter black holes. They find that the entropy quanta depend on the spacetime dimension and are given by $\Delta S = 4 \pi \sin \left( \pi/ (D-1) \right)$. (For large anti-de Sitter black holes see \cite{Wei:2009sw}.) The result for Schwarzschild anti-de Sitter black holes is different from the value (\ref{eq: area entropy quanta}) for the MTBH, although the factor $4 \pi$ appears in both results. We believe that the result by Daghigh and Green \cite{Daghigh:2008jz} deserves further research since it predicts that for the $D$-dimensional Schwarzschild anti-de Sitter black holes their entropy quanta decrease as the spacetime dimension increases.

Thus in Einstein's gravity, the calculated entropy quanta do not depend on the cosmological constant, however Maggiore's method suggests that asymptotically flat and asymptotically anti-de Sitter black holes have different values of the area quanta. (For near extreme Schwarzschild de Sitter black hole see \cite{Li:2009} and for de Sitter spacetime see \cite{LopezOrtega:2009ww}.) Hence Maggiore's proposal predicts that  in Einstein's gravity the entropy quanta are not universal. For other applications of Maggiore's method to get the entropy spectra of black holes in theories of gravity different from general relativity see \cite{Wei:2009yj}, \cite{Fernando:2009tv}, \cite{Kothawala:2008in}, \cite{Setare:2010gi}--\cite{Wei:2010yx}, \cite{Kwon:2010km}, \cite{Gonzalez:2010vv}. 

Making the change of variables \cite{Oliva:2010xn}
\begin{equation}
 r \to \frac{L}{R_+}R, \qquad  \qquad t \to \frac{R_+}{L} \tau,
\end{equation} 
where $R_+$ is a constant, we obtain that the line element (\ref{eq: MTBH metric}) becomes
\begin{equation} \label{eq: ads black hole}
 \dd s^2 = - \frac{(R^2 -R_+^2 )}{L^2} \dd \tau^2 + \frac{L^2 }{(R^2 -R_+^2 )} \dd R^2 + R^2 \dd \Omega_{D-2}^2,    
\end{equation} 
with
\begin{equation}
 \dd \Omega_{D-2}^2 = \frac{L^2}{R_+^2} \dd \Sigma_{D-2}^2 .
\end{equation} 

According to Oliva and Troncoso \cite{Oliva:2010xn} the line element (\ref{eq: ads black hole}) describes an asymptotically anti-de Sitter black hole with an event horizon at $R=R_+$. This black hole is a solution to the equations of motion for several gravity theories as for example, conformal gravity in even dimensions, special cases of Lovelock gravity in odd dimensions, and other gravity theories (see \cite{Klemm:1998kf}--\cite{Aros:2000ij}, \cite{Oliva:2010xn} for more details).

We can suppose that our results on the entropy spectrum of the MTBH with metric (\ref{eq: MTBH metric}) can be extended easily to the black holes with line element (\ref{eq: ads black hole}) since their metrics are similar, but we think that it is necessary to make a careful analysis. We notice that for $D \geq 4$ the QNF of the gravitational perturbations and those of other fields (as for example Klein-Gordon and electromagnetic fields) have a different status, since the quasinormal modes of the gravitational perturbations determine the oscillations in the structure of the spacetime itself, while the quasinormal modes of other fields determine their oscillations in a fixed background. Thus it is not evident that the frequencies of oscillation for the electromagnetic and Klein-Gordon fields can tell something about the spectrum of the event horizon, it is more probable that the information is contained in the oscillation frequencies of the gravitational perturbations. Hence we believe that in Maggiore's method \cite{Maggiore:2007nq} we must use the QNF of the gravitational perturbations to calculate the entropy spectrum.

Although in several spacetimes (as the Schwarzschild black hole, the de Sitter spacetime, and the MTBH) we obtain that the asymptotic physical frequencies of gravitational perturbations and the other fields behave in a similar way, and hence in the entropy spectrum that produces Maggiore's method there is no difference when we use the QNF of a field different from the gravitational perturbations. Thus for $D \geq 4$, when we calculate the entropy spectrum with Maggiore's method and we do not use the QNF of the gravitational perturbations, it is implicitly assumed that the asymptotic physical frequencies of the gravitational perturbations and the other field behave in a similar way. We believe that this hypothesis must be tested for the black holes whose entropy spectra are calculated from the QNF of fields different from the gravitational perturbations in \cite{Setare:2010gi}, \cite{Majhi:2009xh}, \cite{Wei:2010yx}, \cite{Gonzalez:2010vv}.

Since we do not know the QNF of the gravitational perturbations for the black holes (\ref{eq: ads black hole}) when these are solutions of gravity theories different from general relativity and these do not follow in a straightforward way from the QNF of the gravitational perturbations for the MTBH of Einstein's gravity \cite{Birmingham:2006zx}, (we can get the QNF of the Klein-Gordon and electromagnetic fields in the black hole (\ref{eq: ads black hole}) from their QNF in the MTBH \cite{Oliva:2010xn}). We think that in the framework of Maggiore's proposal the calculation of the entropy spectra for the black holes (\ref{eq: ads black hole}) deserves further research.

As we previously mentioned, Gonzalez et al.\ \cite{Gonzalez:2010vv} calculate the entropy spectrum of Chern-Simons black holes. First we note that they use the QNF of the Klein-Gordon field to calculate their entropy spectra. In \cite{Gonzalez:2010vv} the analogous of our adiabatic invariant $I$ are the quantities $I_2$ of formulas (53) and (54). Taking into account their mathematical expressions we find that the two quantities $I_2$ of \cite{Gonzalez:2010vv} go to zero as $M \to 0$ ($\mu \to 0$), and therefore it is not possible to compare the results for the entropy spectra. Also, in \cite{Gonzalez:2010vv} for the Chern-Simons black holes the calculated entropy spectra are not evenly spaced and depend on the properties of the base manifold, in contrast to our result for the MTBH of Einstein's gravity. For the black holes of \cite{Setare:2010gi}, \cite{Majhi:2009xh}, \cite{Wei:2010yx}, \cite{Gonzalez:2010vv} an interesting question is to study whether their entropy spectra change when we use the QNF of the gravitational perturbations to make the computation.

\end{document}